\def\cp{$CP$}
\def\simge{\mathrel{%
   \rlap{\raise 0.511ex \hbox{$>$}}{\lower 0.511ex \hbox{$\sim$}}}}
\def\simle{\mathrel{
   \rlap{\raise 0.511ex \hbox{$<$}}{\lower 0.511ex \hbox{$\sim$}}}}

\documentstyle [11pt]{article}

\title{
\begin{flushright}
{\large UCTP-104-99}
\end{flushright}
\vskip 0.15in
{\bf\boldmath The Time Dependence of $B^0\rightarrow f$ Decays}
}

\author{A.\ J.\ Schwartz \\
{\it Department of Physics, University of Cincinnati, 
Cincinnati, Ohio 45221}}

\date{November 17, 2000}

\begin{document}

\maketitle
\begin{center}
{\bf Abstract}
\end{center}
{\small
We calculate the time dependence of $B^0\rightarrow f$ decays
(and analogously $D^0\rightarrow f$ decays) and use this result to
obtain expressions for the time dependence of: an untagged sample
($B^0\rightarrow f$ and $\overline{B}^0\rightarrow f$ decays
combined), a sample containing charge-conjugate final states
($B^0\rightarrow f$ and $B^0\rightarrow\bar{f}$ decays combined), 
and a sample containing all four decay modes together. For 
simplicity we assume \cp\ violating effects are negligible.
}
\vskip0.35in
The time dependence of $B^0\rightarrow f$ decays, in which $f$ is 
not a \cp\ eigenstate, is not purely exponential due to the presence 
of $B^0$-$\overline{B}^0$ mixing. This mixing arises due to either a
mass difference $\Delta m$ or a decay-width difference $\Delta \gamma$
between the mass eigenstates of the $B^0$-$\overline{B}^0$ system.
This note calculates an expression for the time dependence
in the presence of such mixing. It is assumed that $CP$ violation,
if present, occurs at a negligible level. This assumption is
well-motivated in the $D^0$ system, for which our results 
can be applied. In particular, our results can be applied
to the decay $D^0\rightarrow K^-\pi^+$ \cite{lifediff}.

There are two strong eigenstates, $B^0$ and $\overline{B}^0$, which are
not eigenstates of the full Hamiltonian due to the weak interaction. 
Thus, they most generally evolve according to the Schr\"{o}dinger 
equation:
\begin{equation}
i\frac{\partial\ }{\partial t} \left( \begin{array}{c} | B^0 \rangle \\ 
| \overline{B}^0 \rangle \end{array} \right)
 = \left( {\bf M} - \frac{i}{2}{\bf \Gamma} \right) 
   \left( \begin{array}{c} | B^0 \rangle \\ | \overline{B}^0 \rangle 
\end{array} \right)
\end{equation}
where the $2\times 2$ Hermitian matrices ${\bf M}$ and ${\bf \Gamma}$ 
represent transition amplitudes to virtual (off-mass-shell) and 
real (on-mass-shell) states, respectively. Diagonalizing 
${\bf M} - (i/2){\bf \Gamma}$ yields the physical states
\begin{eqnarray}
| B^{}_S\rangle & \ =\ &  p| B^0 \rangle + q | \overline{B}^0 \rangle 
							\nonumber \\
| B^{}_L\rangle & \ =\ &  p| B^0 \rangle - q | \overline{B}^0 \rangle 
\label{eqn:massdefs}
\end{eqnarray}
where
\begin{equation}
\frac{q}{p} \ =\ \sqrt{\frac{M^*_{12} -(i/2)\,\Gamma^*_{12}}
                            {M^{}_{12}-(i/2)\,\Gamma^{}_{12}}}\ \approx\ 
\sqrt{\frac{M^*_{12}}{M^{}_{12}}} \ =\ e^{i2\phi^{}_{M}}\,.
\label{eqn:qoverpdef}
\end{equation}
In Eq.~(\ref{eqn:massdefs}) we've used the notation $B^{}_L/B^{}_S$
in analogy with the $K^{}_L/K^{}_S$ system,\footnote{Although
$|\tau^{}_{B^{}_L} - \tau^{}_{B^{}_S}| \ll 
                |\tau^{}_{K^{}_L} - \tau^{}_{K^{}_S}|$.}
while in Eq.~(\ref{eqn:qoverpdef}) we've made the good
approximation $\Gamma^{}_{12}\ll M^{}_{12}$ \cite{Gronau}.
The time evolution of the physical states is:
\begin{eqnarray}
| B^{}_S(t)\rangle & \ =\ & |B^{}_S\rangle\,e^{-(\Gamma_S/2 + im^{}_S)t}
                                                              \nonumber \\
| B^{}_L(t)\rangle & \ =\ & |B^{}_L\rangle\,e^{-(\Gamma_L/2 + im^{}_L)t}\,.
\label{eqn:masstime}
\end{eqnarray}
Inverting Eq.~(\ref{eqn:massdefs}) gives:
\begin{eqnarray}
| B^0 \rangle & \ =\ &  
 \frac{1}{2p}\left( | B^{}_S \rangle + | B^{}_L \rangle \right) \nonumber \\
| \overline{B}^0 \rangle & \ =\ &  
 \frac{1}{2q}\left( | B^{}_S \rangle - | B^{}_L \rangle \right)\,,
\label{eqn:neutdefs}
\end{eqnarray}
and inserting the time dependences (\ref{eqn:masstime}) into 
Eqs.~(\ref{eqn:neutdefs}) gives:
\begin{eqnarray}
| B^0(t)\rangle & \ =\ &  
   \frac{1}{2p}\left\{ | B^{}_S \rangle e^{-(\Gamma_S/2 + im^{}_S)t}  
     + | B^{}_L \rangle e^{-(\Gamma_L/2 + im^{}_L)t}  \right\}  \nonumber \\
 & \ =\ &  
e^{-(\overline{\Gamma}/2+i\overline{m})\,t}\ \left\{
   \cosh\left[(\Delta\gamma/4 + i\Delta m/2 )t\right]| B^0 \rangle 
+ \left(\frac{q}{p}\right) 
   \sinh\left[(\Delta\gamma/4 + i\Delta m/2 ) t\right]| \overline{B}^0 \rangle 
                                  \right\}  \nonumber  \\
| \overline{B}^0(t)\rangle & \ =\ &  
 \frac{1}{2q}\left\{ | B^{}_S \rangle e^{-(\Gamma_S/2 + im^{}_S)t}  
       - | B^{}_L \rangle e^{-(\Gamma_L/2 + im^{}_L)t}  \right\} \nonumber \\
 & \ =\ &  
e^{-(\overline{\Gamma}/2+i\overline{m})\,t}\ \left\{ \left(\frac{p}{q}\right)
   \sinh\left[ (\Delta\gamma/4 + i\Delta m/2 )t\right]| B^0 \rangle + 
   \cosh\left[ (\Delta\gamma/4 + i\Delta m/2 )t\right]| \overline{B}^0 \rangle 
  \right\}, \nonumber  
\end{eqnarray}
where $\overline{m}\equiv (m^{}_L + m^{}_S)/2$, 
$\overline{\Gamma}\equiv (\Gamma^{}_L + \Gamma^{}_S)/2$, 
$\Delta m\equiv m^{}_L-m^{}_S$, and  
$\Delta \gamma\equiv \Gamma^{}_L-\Gamma^{}_S$.
In the limit $\Delta\gamma\rightarrow 0$, one recovers
the expressions which follow Eq.~(5) in Ref.~\cite{SchwartzCP}.
The states above lead to decay amplitudes:
\begin{eqnarray}
\langle f|H|B^0(t)\rangle & = & 
e^{-(\overline{\Gamma}/2+i\overline{m})\,t}\ \left\{
   \cosh\left[(\Delta\gamma/4 + i\Delta m/2 )t\right]{\cal A}^{}_f 
+ \left(\frac{q}{p}\right) 
    \sinh\left[(\Delta\gamma/4 + i\Delta m/2 ) t\right]\bar{\cal A}^{}_f 
                                  \right\}  \nonumber  \\
 &  &  \nonumber \\
\langle f|H|\overline{B}^0(t)\rangle & = & 
e^{-(\overline{\Gamma}/2+i\overline{m})\,t}\ \left\{ \left(\frac{p}{q}\right)
    \sinh\left[ (\Delta\gamma/4 + i\Delta m/2 )t\right]{\cal A}^{}_f  + 
   \cosh\left[ (\Delta\gamma/4 + i\Delta m/2 )t\right] \bar{\cal A}^{}_f 
  \right\}, \nonumber  
\end{eqnarray}
where we've defined the amplitudes for 
the pure $B^0$ and $\overline{B}^0$ states as: 
\begin{eqnarray}
{\cal A}^{}_f\ \equiv\ \langle f|H|B^0\rangle & {\rm\ \ \ \ \ \ \ \ \ \ }
\bar{\cal A}^{}_f\ \equiv\ \langle f|H|\overline{B}^0\rangle \ .
\end{eqnarray}
Squaring the decay amplitudes gives the decay rates:
\begin{eqnarray}
|\langle f|H|B^0(t)\rangle |^2 & = &  
   |{\cal A}^{}_f|^2 e^{-\overline{\Gamma} t}\left\{
   |\cosh (\Delta\gamma/4 + i\Delta m/2 )t\ |^2 \ +\  
                                          \right. \nonumber \\
 & &  \hspace*{0.20in} 
     \left|\lambda \right|^2 
        |\sinh (\Delta\gamma/4 + i\Delta m/2 )t\ |^2 \ +\ 
                                                  \nonumber \\
 &  &  \hspace*{0.40in}   \Bigl. 
       \left(\lambda^*\right)   
       \cosh (\Delta\gamma/4 + i\Delta m/2 )t\,
    \left[\sinh (\Delta\gamma/4 + i\Delta m/2 )t\ \right]^*\ +\ c.c.
					\Bigr\}  \nonumber \\
 &  &  \label{eqn:rate1} \\
 &  &  \nonumber \\
|\langle f|H|\overline{B}^0(t)\rangle |^2 & = &  
   |\bar{\cal A}^{}_f|^2 e^{-\overline{\Gamma} t}\left\{
   |\cosh (\Delta\gamma/4 + i\Delta m/2 )t\ |^2 \ +\  
                                          \right. \nonumber \\
 & &  \hspace*{0.20in} 
     \left|\bar{\lambda} \right|^2 
        |\sinh (\Delta\gamma/4 + i\Delta m/2 )t\ |^2 \ +\ 
                                                  \nonumber \\
 &  &  \hspace*{0.40in}   \Bigl. 
    \left(\bar{\lambda}^*\right)   
       \cosh (\Delta\gamma/4 + i\Delta m/2 )t\,
      \left[\sinh (\Delta\gamma/4 + i\Delta m/2 )t\ \right]^*\ +\ c.c.
					\Bigr\}\,, \nonumber  \\
 &  &  \label{eqn:rate2}
\end{eqnarray}
where we've defined the parameters
$\lambda\equiv (q/p)(\bar{\cal A}^{}_f/{\cal A}^{}_f)$ and
$\bar{\lambda}\equiv (p/q)({\cal A}^{}_f/\bar{\cal A}^{}_f)$. 
To evaluate these expressions, note that:
\begin{eqnarray}
|\cosh (\Delta\gamma/4 + i\Delta m/2 )t |^2 & = & 
|\cosh (\Delta\gamma/4)t\,\cos (\Delta m/2)t + 
   i\sinh (\Delta\gamma/4)t\,\sin (\Delta m/2)t |^2  \nonumber \\
 & = &   \cosh^2(\Delta\gamma/4)t\,\cos^2(\Delta m/2)t + 
    \sinh^2(\Delta\gamma/4)t\,\sin^2(\Delta m/2)t   \nonumber \\
 & = &   \cosh^2(\Delta\gamma/4)t - \sin^2(\Delta m/2)t   \nonumber \\
 & = &  \left(\frac{1}{2}\right)\Bigl[ \cosh (\Delta\gamma/2)t + 
                        \cos (\Delta m)t \Bigr]\,.
\label{eqn:costerm}
\end{eqnarray}
Similarly,
\begin{eqnarray}
|\sinh  (\Delta\gamma/4 + i\Delta m/2 )t |^2 & = & 
|\sinh (\Delta\gamma/4)t\,\cos (\Delta m/2)t + 
   i\cosh (\Delta\gamma/4)t\,\sin (\Delta m/2)t |^2  \nonumber \\
 & = &   \sinh^2(\Delta\gamma/4)t\,\cos^2(\Delta m/2)t + 
    \cosh^2(\Delta\gamma/4)t\,\sin^2(\Delta m/2)t   \nonumber \\
 & = &   \cosh^2(\Delta\gamma/4)t - \cos^2(\Delta m/2)t   \nonumber \\
 & = &  \left(\frac{1}{2}\right)\Bigl[ \cosh (\Delta\gamma/2)t - 
                        \cos (\Delta m)t \Bigr]\,.
\label{eqn:sinterm}
\end{eqnarray}
Finally, 
\begin{eqnarray}
 &  &  \cosh (\Delta\gamma/4 + i\Delta m/2 )t\,
      \left[\sinh (\Delta\gamma/4 + i\Delta m/2 )t\right]^* \ =\
						\nonumber \\
 &  &  \nonumber \\ 
 &  & \hspace*{0.20in}
       \left[\cosh (\Delta\gamma/4)t\,\cos (\Delta m/2)t + 
   i\sinh (\Delta\gamma/4)t\,\sin (\Delta m/2)t \right] \ \times\ 
      \nonumber \\
 &  & \hspace*{0.80in}
 \left[\sinh (\Delta\gamma/4)t\,\cos (\Delta m/2)t - 
   i\cosh (\Delta\gamma/4)t\,\sin (\Delta m/2)t \right] \nonumber \\
 &  &  \nonumber \\ 
 & = & \hspace*{0.20in}
\left(\frac{1}{2}\right)
\left[\sinh (\Delta\gamma/2)t\,\cos^2(\Delta m/2)t + 
      \sinh (\Delta\gamma/2)t\,\sin^2(\Delta m/2)t\ +\ 
           \right. \nonumber \\
 &  &     \hspace*{0.80in}
   \left. i\sinh^2(\Delta\gamma/4)t \sin (\Delta m)t -
   i\cosh^2(\Delta\gamma/4)t \sin (\Delta m)t\right]  \nonumber \\
 & = & \hspace*{0.20in}
\left(\frac{1}{2}\right) 
\Bigl[\sinh (\Delta\gamma/2)t - i\sin (\Delta m)t \Bigr]\,. 
\label{eqn:crossterm}
\end{eqnarray}
\vskip0.15in\noindent
Inserting Eqs.~(\ref{eqn:costerm}), (\ref{eqn:sinterm}), and 
(\ref{eqn:crossterm}) into Eqs.~(\ref{eqn:rate1}) and 
(\ref{eqn:rate2}) gives:
\begin{eqnarray} 
|\langle f|H|B^0(t)\rangle |^2 & = &  
   \left(\frac{|{\cal A}^{}_f|^2}{2}\right) e^{-\overline{\Gamma} t}
\Bigl\{\left[ \cosh (\Delta\gamma/2)t + \cos (\Delta m)t \right]
				\ +\  \!\Bigr. \nonumber \\
 & &  \hspace*{0.6in} 
     \left|\lambda \right|^2 
\left[ \cosh (\Delta\gamma/2)t - 
                        \cos (\Delta m)t \right] \ +\ 
                                  \nonumber \\
 &  &  \hspace*{0.80in}   
  \left(\lambda^*\right)\left[
     \sinh (\Delta\gamma/2)t\ - i\sin (\Delta m)t\ \right]  \ +\
        \nonumber \\
 &  &  \hspace*{1.2in}   \Bigl. 
   \left(\lambda\right)\left[ 
     \sinh (\Delta\gamma/2)t\ + i\sin (\Delta m)t\ \right] \!\Bigr\}  
\nonumber \\
 & = & 
   \left(\frac{|{\cal A}^{}_f|^2}{2}\right) e^{-\overline{\Gamma} t}
\Bigl\{ (1+|\lambda|^2)\cosh (\Delta\gamma/2)t \ + 
(1-|\lambda|^2)\cos (\Delta m)t  \ +\ \!\Bigr. \nonumber \\
 &   & \hspace*{1.4in} \Bigl.  
(\lambda + \lambda^*)\sinh (\Delta\gamma/2)t\ + 
i(\lambda - \lambda^*)\sin (\Delta m)t\ \!\Bigr\} \nonumber \\
  &  &  \label{eqn:result1}  \\
 &  &  \nonumber \\
|\langle f|H|\overline{B}^0(t)\rangle |^2 & = &  
   \left(\frac{|\bar{\cal A}^{}_f|^2}{2}\right) e^{-\overline{\Gamma} t}
\Bigl\{ \left[ \cosh (\Delta\gamma/2)t + \cos (\Delta m)t \right]
				\ +\  \!\Bigr. \nonumber \\
 & &  \hspace*{0.6in} 
     \left|\bar{\lambda} \right|^2 
\left[ \cosh (\Delta\gamma/2)t - 
                        \cos (\Delta m)t \right] \ +\ 
                                  \nonumber \\
 &  &  \hspace*{0.80in}   
  \left(\bar{\lambda}^*\right)\left[
     \sinh (\Delta\gamma/2)t\ - i\sin (\Delta m)t\ \right]  \ +\
        \nonumber \\
 &  &  \hspace*{1.2in}   \Bigl. 
   \left(\bar{\lambda}\right)\left[ 
     \sinh (\Delta\gamma/2)t\ + i\sin (\Delta m)t\ \right] \!\Bigr\}  
\nonumber \\
 & = & 
   \left(\frac{|\bar{\cal A}^{}_f|^2}{2}\right) e^{-\overline{\Gamma} t}
\Bigl\{ (1+|\bar{\lambda}|^2)\cosh (\Delta\gamma/2)t \ + 
(1-|\bar{\lambda}|^2)\cos (\Delta m)t  \ +\ \!\Bigr. \nonumber \\
 &   & \hspace*{1.4in} \Bigl.  
(\bar{\lambda} + \bar{\lambda}^*)\sinh (\Delta\gamma/2)t\ + 
i(\bar{\lambda} - \bar{\lambda}^*)\sin (\Delta m)t\ \!\Bigr\}\,.\nonumber \\
  &   & \label{eqn:result2}  
\end{eqnarray}
\vskip 0.15in
If the lifetime distribution for a final state $f$ is constructed
from a sample of {\it untagged\/} decays, and there are equal numbers 
of $B^0$ and $\overline{B}^0$ mesons produced, then the time dependence
of the decays will be the sum of Eqs.~(\ref{eqn:result1}) and
(\ref{eqn:result2}). If $|\bar{\cal A}^{}_f|\ll |{\cal A}^{}_f|$
(e.g., if $B^0\rightarrow f$ is Cabibbo-favored and  
$\overline{B}^0\rightarrow f$ is doubly-Cabibbo-suppressed), 
then one would measure:
\begin{eqnarray}
\frac{dN^{}_{(B^0+\overline{B}^0)\rightarrow f}}{dt} & \approx & 
   \left(\frac{1}{2}\right) e^{-\overline{\Gamma} t}\left\{
2|{\cal A}^{}_f|^2 \cosh (\Delta\gamma/2)t \ + 
\right. \nonumber \\
 &  &  \hspace*{0.2in} 
{\cal A}^*_f \bar{\cal A}^{}_f \left[ (q/p) + (p/q)^*\right] 
				\sinh (\Delta\gamma/2)t\  +
{\cal A}^{}_f \bar{\cal A}^*_f \left[ (q/p)^* + (p/q)\right] 
				\sinh (\Delta\gamma/2)t\  + 
        \nonumber \\
 &  & \hspace*{0.5in} \left.
{\cal A}^*_f \bar{\cal A}^{}_f \left[ (q/p) - (p/q)^*\right] 
				i\sin (\Delta m)t\  +
{\cal A}^{}_f \bar{\cal A}^*_f \left[ (p/q) - (q/p)^*\right] 
				i\sin (\Delta m)t\ \right\} \nonumber \\
 & = & 
   \left(\frac{1}{2}\right) e^{-\overline{\Gamma} t}\left\{
2|{\cal A}^{}_f|^2 \cosh (\Delta\gamma/2)t \ + 
\left[ {\cal A}^*_f \bar{\cal A}^{}_f (2q/p) +
{\cal A}^{}_f \bar{\cal A}^*_f (2q/p)^* \right] 
			\sinh (\Delta\gamma/2)t\ \right\} \nonumber \\
 & = & 
  |{\cal A}^{}_f|^2 e^{-\overline{\Gamma} t}\Bigl\{
 \cosh (\Delta\gamma/2)t \ + (\lambda + \lambda^*)\sinh (\Delta\gamma/2)t\ 
				\!\Bigr\}\,,
\label{eqn:untagged}  
\end{eqnarray}  
where we've used the fact that $|q/p|^2 = 1$.
This result [and Eq.~(\ref{eqn:final0}) below]
was previously obtained by Dunietz \cite{Dunietz}.

If the lifetime distribution includes final states $f$ and
$\bar{f}$ combined together, then we must also consider the decay
rates $|\langle \bar{f}|H|\overline{B}^0(t)\rangle |^2$ and  
$|\langle \bar{f}|H|B^0(t)\rangle |^2$. These rates are
equivalent to Eqs.~(\ref{eqn:result1}) and (\ref{eqn:result2}),
respectively, with the interchange $q/p \leftrightarrow p/q$
(since $|{\cal A}^{}_f| = |\bar{\cal A}^{}_{\bar{f}}|$ and
$|\bar{\cal A}^{}_f| = |{\cal A}^{}_{\bar{f}}|$ by $CP$ conservation).
The analog of Eq.~(\ref{eqn:untagged}) for the final state $\bar{f}$ 
is then:
\begin{eqnarray}
\frac{dN^{}_{(B^0+\overline{B}^0)\rightarrow\bar{f}}}{dt} & \approx & 
  |{\cal A}^{}_f|^2 e^{-\overline{\Gamma} t}\Bigl\{
 \cosh (\Delta\gamma/2)t \ + (\kappa + \kappa^*)\sinh (\Delta\gamma/2)t\ 
				\!\Bigr\}\,,   \label{eqn:final0}
\end{eqnarray}
where $\kappa\equiv (p/q)({\cal A}^{}_{\bar{f}}/\bar{\cal A}^{}_{\bar{f}})
= (q/p)^*(\bar{\cal A}^{}_f/{\cal A}^{}_f)e^{i\theta}$.
Thus, 
\begin{eqnarray}
\frac{dN^{}_{(B^0+\overline{B}^0)\rightarrow (f+\bar{f})}}{dt} & \approx & 
  2|{\cal A}^{}_f|^2 e^{-\overline{\Gamma} t}\Bigl\{
 \cosh (\Delta\gamma/2)t \ + Re (\lambda + \kappa)\sinh (\Delta\gamma/2)t\ 
				\!\Bigr\}\,. \nonumber \\
 &  & \label{eqn:final1}
\end{eqnarray}
For this expression to be valid the $f$ and $\bar{f}$ final
states must have equal acceptances, or else the decays must be
corrected for acceptance before the lifetime distributions are 
combined.

If the $B^0$ and $\overline{B}^0$ mesons are produced in 
a fixed-target experiment, then their yields will usually
be different due to there being unequal numbers of $d$ and $\bar{d}$ 
quarks in the initial state. In this case Eq.~(\ref{eqn:untagged}) 
does not apply, and it is more convenient to consider 
$B^0$ and $\overline{B}^0$ samples separately. Such
experiments typically combine together the final states 
$f$ and $\bar{f}$ (when not studying \cp\ asymmetries), 
and in this case:
\begin{eqnarray}
\frac{dN^{}_{B^0\rightarrow (f+\bar{f})}}{dt} & = & 
  |{\cal A}^{}_f|^2 e^{-\overline{\Gamma} t}\Bigl\{
\left( 1+|\lambda|^2\right) \cosh (\Delta\gamma/2)t \ + \Bigr. 
						\label{eqn:final2} \\
 & & \hspace*{0.50in} \Bigl.
Re (\lambda + \kappa^*)\sinh (\Delta\gamma/2)t \ -
   Im (\lambda + \kappa^*)\sin (\Delta m)t  \Bigr\} \nonumber \\
 &  &  \nonumber \\
\frac{dN^{}_{\overline{B}^0\rightarrow (f+\bar{f})}}{dt} & = & 
  |{\cal A}^{}_f|^2 e^{-\overline{\Gamma} t}\Bigl\{
\left( 1+|\lambda|^2\right) \cosh (\Delta\gamma/2)t \ + \Bigr. 
						\label{eqn:final3} \\
 & & \hspace*{0.50in} \Bigl.
Re (\lambda^* + \kappa)\sinh (\Delta\gamma/2)t \ -
   Im (\lambda^* + \kappa)\sin (\Delta m)t  \Bigr\}\,. \nonumber 
\end{eqnarray}
Adding together Eqs.~(\ref{eqn:final2}) and (\ref{eqn:final3}) 
(corresponding to an untagged sample having equal numbers of 
$B^0$ and $\overline{B}^0$) recovers Eq.~(\ref{eqn:final1}).


\begin{thebibliography}{99}

\bibitem{lifediff} E.\ M.\ Aitala {\it et al.\/} (E791 Collaboration),
		``Measurements of Lifetimes and a Limit on the Lifetime
		Difference in the Neutral $D$-Meson System,''
		Phys.\ Rev.\ Lett.\ 83 (1999) 32.

		J.\ M.\ Link {\it et al.\/} (FOCUS Collaboration),
		``A Measurement of Lifetime Differences in the
		Neutral $D$-Meson System,''
		Phys.\ Lett.\ B485 (2000) 62.

\bibitem{Gronau} See, e.g.: M.\ Gronau, in {\it Proceedings of the 
		 Workshop on B Physics at Hadron Accelerators}, Snowmass,
                 Colorado, ed.\ P.\ McBride and C.\,S.\,Mishra,
                 Fermilab-CONF-93/267 (1993).

\bibitem{SchwartzCP} A.\ J.\ Schwartz, ``General Formalism of $CP$
		Violation in $B^0$ Decays,'' PRINCETON/HEP/96-09
		(1996) (unpublished).

\bibitem{Dunietz} I.\ Dunietz, ``$B^{}_s$-$\overline{B}^{}_s$ Mixing,
		  $CP$ Violation, and Extraction of CKM Phases from
		  Untagged $B^{}_s$ Data Samples,'' 
		  Phys.\ Rev.\ D52 (1995) 3048.

\end{thebibliography}
\end{document}